# The Anatomy of Big Data Computing

Raghavendra Kune[1], Pramod Kumar Konugurthi[1], Arun Agarwal[2], Raghavendra Rao Chillarige[2], and Rajkumar Buyya[3]

[1]Advanced Data Processing Research Institute, Department of Space, India
[2] School of Computer and Information Sciences, University of Hyderabad, India
[3]CLOUDS Lab, Department of Computing and Information Systems, University of Melbourne, Australia

{raghav.es, pramodkumar.konugurthi, aruncs.2011}@gmail.com, vijaya_crr@yahoo.co.in, rbuyya@unimelb.edu.au

*Abstract- Advances in information technology and its widespread growth in several areas of business, engineering, medical and scientific studies are resulting in information/data explosion. Knowledge discovery and decision making from such rapidly growing voluminous data is a challenging task in terms of data organization and processing, which is an emerging trend known as **Big Data Computing**; a new paradigm which combines large scale compute, new data intensive techniques and mathematical models to build data analytics. Big Data computing demands a huge storage and computing for data curation and processing that could be delivered from on-premise or clouds infrastructures. This paper discusses the evolution of Big Data computing, differences between traditional data warehousing and Big Data, taxonomy of Big Data computing and underpinning technologies, integrated platform of Big Data and Clouds known as Big Data Clouds, layered architecture and components of Big Data Cloud and finally discusses open technical challenges and future directions.*

## 1 Introduction

Big Data computing is an emerging data science paradigm of multi dimensional information mining for scientific discovery and business analytics over large scale infrastructure. The data collected/produced from several scientific explorations and business transactions often require tools to facilitate efficient data management, analysis, validation, visualization and dissemination, while preserving the intrinsic value of the data [1] - [5]. The IDC [6] report predicted that there could be an increase of the digital data by 40 times from 2012 to 2020. New advancements in semiconductor technologies are eventually leading to faster computing, large scale storage, faster and powerful networks at lower prices, enabling large volumes of data preservation and utilization at faster rate. Recent advancements in Cloud computing technologies are enabling to preserve, every bit of the gathered and processed data, based on subscription models, providing high availability of storage and computation at affordable price. Conventional data warehousing systems are based on pre-determined analytics over the abstracted data, and employs cleansing and transforming into another database known as data marts- which are periodically updated with the similar type of rolled-up data. However, Big Data systems work on non predetermined analytics; hence no need of data cleansing and transformations procedures.

Big Data organizes and extracts the valued information from the rapidly growing, large volumes, variety forms, and frequently changing data sets collected from multiple, and autonomous sources in the minimal possible time, using several statistical, and machine learning techniques. Big Data is characterized by 5V's such as Volume, Velocity, Variety, Veracity, and Value. Big Data and traditional data warehousing systems, however, have the similar goals to deliver business value through the analysis of data, but, they differ in the analytics methods and the organization of the data. In practice, data warehouses organize the data in the repository, by collecting it from other several databases like enterprise's financial systems, customer marketing systems, billing systems, point-of-sale systems, etc.

Warehousing systems are poor on organizing and querying the data from the operational streaming data like click stream logs, sensor data, location data from mobile devices, customer support emails and chat transcripts, and surveillance videos etc. Big Data technologies overcome the weakness of the data warehousing systems, by harnessing new sources of data, thus facilitating enterprises analyze and extract intrinsic information through analytics. Big Data technology has gaining popularity in several domains of business, engineering and scientific computing areas, Philip et.al [8] presented a survey on Big Data along with opportunities and challenges for data intensive applications stated several areas and the importance of Big Data. M. Chen et.al [9] presented a survey on Big Data and its interrelated technologies like Clouds, Internet of Things (IOT), online social networks, medical applications, collective intelligence, and smart grid. J. Chen et.al [10] presented the Big Data technologies towards data management challenges like big data diversity, big data reduction, integration and cleaning, indexing and query, and several tools for analysis and mining. X. Wu et.al [11] presented Big Data processing model, from the data mining perspective. Kaiser et.al [12] discussed several issues in Big Data such as storage and data transport technologies followed by methodologies for Big Data Analytics. R. Buyya et.al [13] presented a survey on Big Data computing in Clouds and future research directions for the development of analytics and visualization tools in several domains of science, engineering and business.

As business domains are growing, there is a need to converge a new economic system redefining the relationships among producers, distributors and consumers of goods and services. Obviously, it is not feasible to depend on experience or pure intuition always, however, it is also essential to use critically important data sources for decision making. NIST Big Data Public Working Group described a survey of Big Data Architectures and Framework from the industry [14] . The several areas of Big Data computing are described below.

**a) Scientific Explorations:** The data collected from various sensors is analyzed to extract the useful information for societal benefits. E.g. physics and astronomical experiments- a large number of scientists collaborating for designing, operating and analyzing the products of sensor networks and detectors for scientific studies. Earth Observation Systems (EOS) - information gathering and analytical approaches about earth's physical, chemical and biological systems via remote sensing technologies, to improve social and economic well-being and its applications for weather forecasting, monitoring and responding to natural disasters, and climate change predictions etc.

**b) Healthcare:** Healthcare organizations would like to predict the locations from where the diseases are spreading so as to prevent further spreading [15] . However, to predict exactly the origin of the disease would not be possible, until there is statistical data from several locations. In 2009, when a new flu virus similar to H1N1 was spreading, Google has predicted this and published a paper in the scientific journal *Nature* [16], by looking at what people were searching for, on the internet.

**c) Governance:** Surveillance system analyzing and classifying streaming acoustic signals, transportation departments using real-time traffic data to predict traffic patterns, update public transportation schedules. Security departments analyzing images from aerial cameras, news feeds, and social networks or items of interest. Social program agencies gain a clearer understanding of beneficiaries and proper payments. Tax agencies identifying fraudsters and support investigation by analyzing complex identity information and tax returns. Sensor applications such stream air, water and temperature data to support cleanup, fire prevention and other programs.

**d) Financial and Business Analytics:** Retaining customers and satisfying consumer expectations are among the most serious challenges facing financial institutions. Sentiment analysis and predictive analysis would play a key role in several fields like travel industry- for optimal cost estimations, retail industry- products targeted for potential customers, Forecast analysis – estimating the best price estimations etc.

**e) Web Analytics:** Several web sites are experiencing millions of unique visitors per day, in turn creating a large range of content. Increasingly, companies want to be able mine this data to understand limitations of their sites, improve response time, offer more targeted ads and so on. This requires tools to perform complicated analytics on data that far exceeds the memory of a single machine or even in cluster of machines.

Service oriented technologies aka Cloud computing are delivering compute, storage and software applications as services over private or public networks based on pay-as-go delivery models [17] [18]. Cloud computing technologies becoming a reality, it is serving as a key enabler for Big Data to solve data intensive problems over a large scale infrastructure for information extraction. The integration of Big Data technologies and Cloud computing read as - "Big Data Clouds" is an emerging new generation data analytics platform for information mining, knowledge discovery and decision making. Hence, both the technologies put together, here, we discuss the evolution of Big Data technologies and compare it with traditional data warehousing technologies along with its relationship with Cloud computing technologies and infrastructure. We also discuss the architecture and reference framework for Big Data computing on Clouds. This paper is intended for researchers, technical audience of both developer and designers, and general readers who are interested in acquiring an in-depth knowledge of Big Data technology in IT.

The rest of the paper is organized as follows. Section 2 describes the differences between Big Data and traditional data ware housing systems in, data handling, processing, storing, extracting etc. followed by CAP theorem- the fundamental principle of database system, and illustrate the ACID and BASE properties adopted for data warehousing (relational model) and Big Data models respectively. Later, we discuss the Big Data abstraction layers and compare it with the traditional data base model. Section 3 discusses taxonomy of Big Data computing, and presents a detailed study of several components of the taxonomy like analytics, frameworks, technologies, programming models, schedulers, processing tools etc. Section 4 illustrates an integrated platform for Big Data and Clouds followed by a layered architecture and its components. Here, we discuss elements of Big Data Cloud, Layered architecture and reference framework. Section 5 identifies open technical challenges, gap analysis and future research directions in data storage/handling, specific domain areas, new programming models and domain specific analytics development.

## 2 Big Data Characteristics – Traditional Data Vs Big Data paradigms

Big Data refers large scale data architectures, and facilitate tools addressing new requirements in handling data volume, velocity, and variability. Traditional databases (data warehousing) assume data is organized in rows and columns and employs data cleansing methods on the data while the data volumes grow over a time period, and often lack on handling such large scale data processing. Traditional Data base / warehousing systems were designed to address smaller volumes of structure data, with the predictable updates and consistent data structure, that mostly operate on single server and lead to operational expenses with the increased data volume. However, Big Data comes in a variety of diverse formats with both batch and stream processing in several areas such as geospatial data, 3D data, audio and video, structured data, unstructured text including log files , sensor data and social media. Below, we discuss the properties of traditional database (Data Warehousing) and Big Data.

## 2.1 Traditional Database (both Operational OLTP and Warehousing OLAP data)

Bill Inmo [19] described data warehousing as subject oriented, integrated, time-variant, and nonvolatile collection of data, and helping analysts in decision making process. Data warehouse is segregated from the organization's operational database. The operational database undergoes the per day transactions (On Line Transaction Processing – OLTP) which causes the frequent changes to the data on daily basis. Traditional databases typically addresses the applications for business intelligence, however, lack in providing the solutions for unstructured large volumes rapidly changing analytics in business and scientific computing. The several processing techniques under data warehouse are described below.
- o analytical processing involves analyzing the data by means of basic OLAP (Online Analytical Processing) operations, including slice-and-dice, drill down, drill up, and pivoting.
- o knowledge discovery through mining techniques by finding the pattern and associations, constructing analytical models, performing classification and prediction. These mining results can be presented using visualization tools.

## 2.2 Big Database:
Big Data addresses the data management and analysis issues in several areas of business intelligence, engineering and scientific explorations. Traditional databases segregate the operational and historical data for operational and analysis reasoning, which are mostly structured. However, Big Data bases address the data analytics over an integrated scale out compute and data platform for unstructured data in near real time. Figure 1 depicts several issues in Traditional data (Data warehousing OLTP/OLAP) and Big Data technologies which are classified into major areas like infrastructure, data handling and decision support software as described below.
- *Decision support / intelligent software tools*: Big Data technologies addresses various decision supporting tools for searching the large data volumes and constructs the relations and extract the information based on several analytical methods. These tools would address several machine learning techniques, decision support systems and statistical modeling tools.
- *Large scale data handling:* rapidly growing data distributed over several storages and compute nodes with multi-dimensional data formats.
- *Large scale infrastructure:* scale out infrastructure for efficient storage and processing.
- *Batch and stream support:* capability to handle both batch and stream computation.

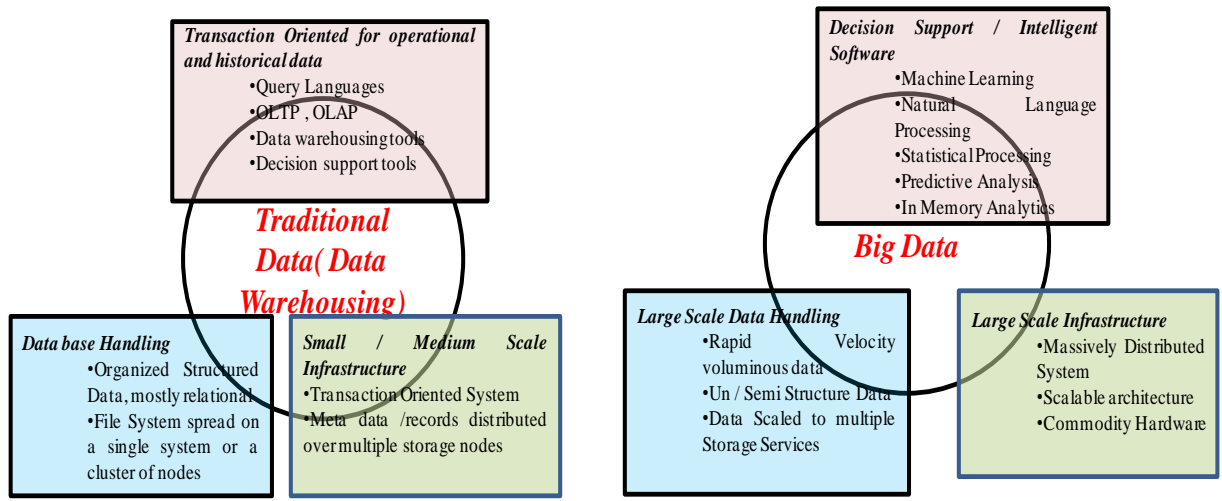

**Figure 1. Big Data Vs Traditional Data (Data Warehousing) models**

Table 1 illustrates properties of Big Data Vs Traditional Data Warehousing computing.

**Table 1. Traditional Data warehousing Vs Big Data issues**

| S.no | Property | Traditional Data Warehousing | Big Data specific issues |
|---|---|---|---|
| 1 | data volume | Data is segregated into operational and historical data. Applies ETL (Extract, Transformation and Load) mechanisms for processing. As the data volumes are increased, the historical data is filtered from ware house system for faster database queries. | High volume of data from several sources like web, sensor networks, social networks, scientific experiments. Capable of handling operational and historical data together, which could be replicated on multiple storage devices for high availability and throughput. |
| 2 | speed | Transaction oriented and the data in turn generated from the transactions is low. | High data growth due to several sources like web and scientific sensors streaming experiments. |
| 3 | data formats | Semi/structured data like XML and Relational. | Multi structured data handling such as relational, and un/semi structured such as text, XML, video streaming etc. |
| 4 | applicable platforms | OLTP (Online Transactional Processing), Relational RDBMS. | Big data analytics, text mining, video analytics, web log mining, scientific data exploration, and intrinsic information extractions, graph analytics, social networking, In memory analytics, statistical and predictive analytics. |
| 5 | programming methodologies/ languages | Query Language like SQL. | Data intensive computing languages for batch processing and stream computing like map/reduce, |

| | | | NoSQL programming. |
|---|---|---|---|
| 6 | data backup / archival | Files / relational data need to have data backup procedures or mechanisms. Traditional data works on regular, incremental and full backup mechanisms that are already established. | Due to large and high speeds of the data growth rates, the conventional methods are not adequate; hence techniques such as differential backup mechanisms need to be explored. |
| 7 | disaster recovery (DR) | Data is replicated at several places to address the disaster. | DR techniques could be separated from mission critical and non critical data. |
| 8 | relationship with Clouds | Relational Data bases/ Data ware housing tools as services over cloud infrastructures. | On demand Big Data infrastructure setup, analytic services by several cloud and Big Data providers. |
| 9 | data de-duplication | Applicable to transactional record deduplication while merging database records. | File and block level deduplication mechanisms need to be explored for continuous growing and stream oriented data. |
| 10 | System users | Administrators, developers and end users. | Data scientists, analytics end users. |
| 11 | theorem applicable | Follows CAP theorem [20] with ACID [21] properties. | Follows CAP theorem with BASE properties [22]. |

### 2.3 *CAP Theorem – ACID and BASE*

Traditional databases follow ACID [21] [23] properties, which are the primary standards for relational databases. However, distributed computing systems follow BASE [22] properties to address loss of consistency and reliability as discussed below:

- **Basically available**: This property states that, the system guarantees the data availability, however, during the transition/changing state the response would be either delayed or may fail in obtaining the requested data. This scenario, is similar as depositing a check in your bank account, and waiting till the check goes through the clearing house, for having the funds made available.
- **Soft state**: The state of the system would change over time, so even during times without input, there may be changes going on due to eventual consistency, thus state of the system is always soft.
- **Eventual consistency**: The system would propagate the data as it is receiving, however, will not ensure the consistency of the data for every transaction. The data would be eventually consistent, whenever it stops receiving the input.

In 2000, Eric Brewer presented CAP theorem, also known as Brewer's theorem [20] for the successful design, implementation and deployment of applications in distributed computing systems. CAP theorem states that any networked shared-data system can provide only two out of the following three properties mentioned below.

- **Consistency**: similar to the consistency property of ACID, the data is synchronized across all cluster nodes, and all the nodes would see the similar data at the same time.
- **Availability**: guaranteed that every request receives a response however, the request is successful/failed in receiving the data which has been requested would not be known.

• **Partition tolerance**: single node failure should not cause the entire system to fail and the system should continue to function even under circumstances of arbitrary message loss or partial failure of the system.

Big Data system adopts Brewer's CAP theorem on the similar lines of BASE. CAP theorem with ACID and BASE is depicted in Figure 2.

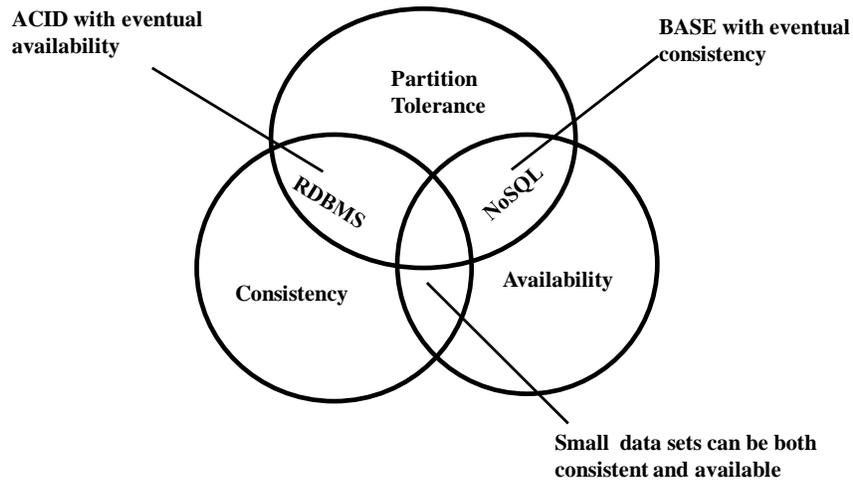

**Figure 2. CAP theorem with ACID and BASE (Source: NIST [24]).**

## 2.4 *Big Data – Abstraction Layers*

Big Data and the traditional/relational data layers are depicted in Figure 3.

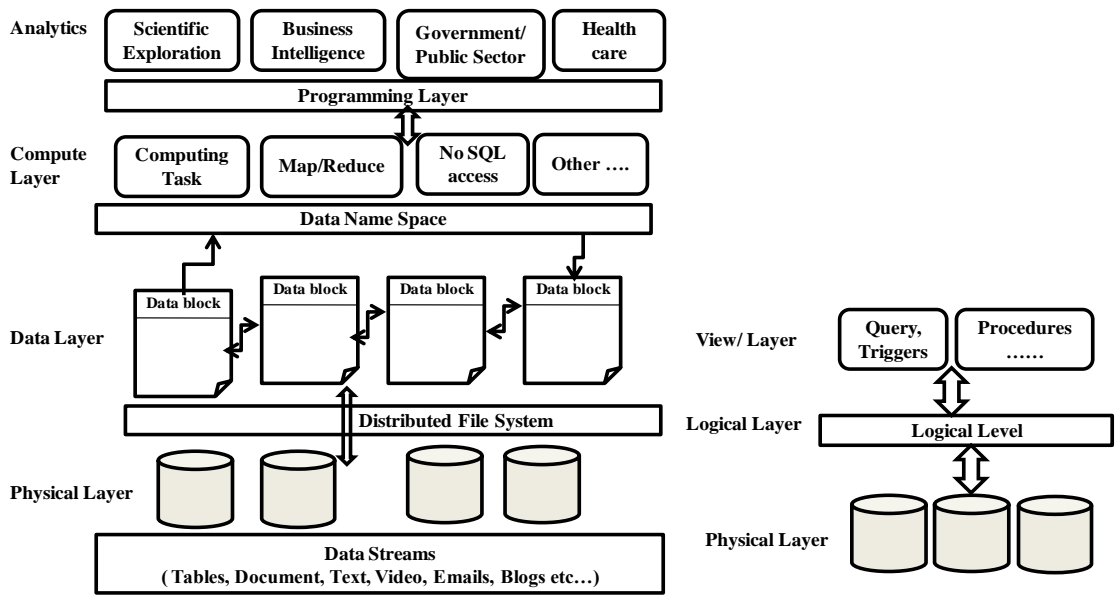

**Figure 3.1. Big Data Abstraction.     Figure 3.2. Traditional Data Abstraction.**

**Figure 3. Big Data Vs Traditional Data Abstraction**

Traditional data layers are in general classified into three layers of abstraction; physical layer, logical layer and view/user layer. The function of each layer is described below.

- Physical layer: It describes the lowest level of abstraction, and uses low level complex data structures for data storage, for example B+-tree organization, R-Tree indexing mechanisms etc.
- Logical layer: describes type of data stored in a database, and the relationships among the data. In general, data base administrators (DBAs) work at the logical level of abstraction. The several activities by this layer are data base design, tuning for better performance, tools for data base backup etc.
- View layer: Highest level of abstraction, hides all low level complexities, details of the data types, and offers programming tools for query and processing.

Big Data abstraction adopts four layered abstraction model, the layers from the bottom up approach are physical layer, data layer, computing layer and data analytics layers respectively. Physical layer takes care of the data organization over several distributed data storage, high speed networks, and partitioned cluster of nodes. Data layer addresses the global namespace for the data access and logical expansion of the data, without knowing the underlying physical layer structure. Compute layer offers several computing methodologies and analytic offers several technologies for analysis of the data for decision making. The role of each layer is described below.

- Physical data layer: Big Data addresses several forms/types of data with a horizontally scalable infrastructure for redundancy, high performance data transfer and efficient support for computation. This layer addresses the properties mentioned against serial numbers 1, 2 and 3 in Table 1.
- Data layer: This layer provides an abstraction over the physical data layer and offers the core functionalities such as data organization, access and retrieval of the data. This layer indexes the data and organizes them over the distributed store devices. The data is partitioned into blocks and organized onto the multiple repositories. The data to organize can be any one of several forms; hence, several tools and techniques are employed for effective organization and retrieval of the data. The examples include key/value pair; column oriented data, document database, Relational database, semi structured XML data, raw formats etc. This layer refers to the properties 3, 6 and 10 mentioned in the Table 1.
- Computing layer: software abstraction layer of data modeling and query, domain specific programming APIs to retrieve the data from its below data model layer. For example, this layer offers tools like NoSQL programming, and Map Reduce for data intensive computing, domain specific statistical models, machine learning techniques etc. This layer refers to the properties 7, 9 and 16 of the Table 1.
- Analytics: standards and techniques for developing the domain specific analytics tools using the tools of software abstraction layer.

## 3  Big Data Taxonomy

For years, several organizations are capturing the transactional structured data using traditional relational data bases using transactional query processing [1] [6] [25] for the information extraction. In recent years, technologies are evolving to perform the investigations on the whole data using distributed computing and storage technologies such as map reduce, distributed file systems and in-memory computing [26] with highly optimized capabilities for different business and scientific purposes. The advancements in storage capacity, data handling and processing tools, the analysis of data can be carried out in real time or close to real time, acting on full data sets rather than on the summarized elements, leveraging tools and technologies enough to address the issue. In addition, the number of options to interpret and analyze the data has also increased, with the use of various visualization technologies. Below, we describe the taxonomy of Big Data depicted in Figure 4. The several elements of the taxonomy are described below.

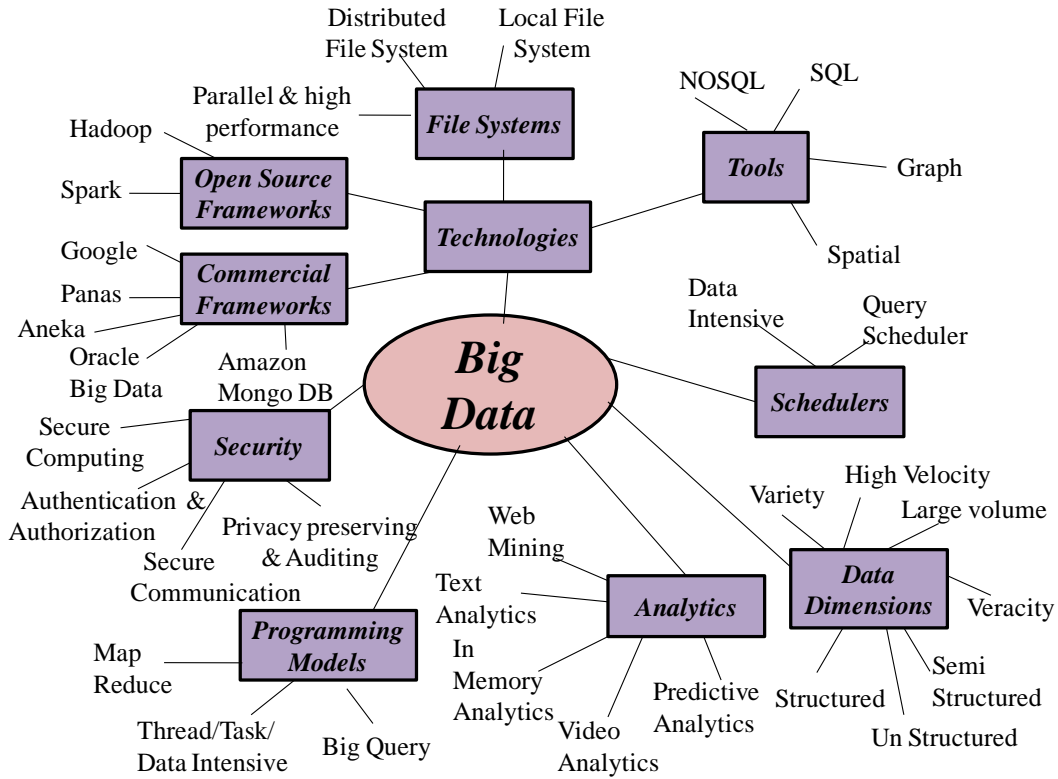

**Figure 4. Big Data Taxonomy**

### i) *Big Data Dimensions*

Big Data is characterized into four dimensions called 4V's; Volume, Velocity, Variety, Veracity as depicted in Figure 5. Aside, another dimension V (Value/Valor) also used to characterize the quality of the data.

- **Volume:** Volume is concerned about scale of data i.e. the volume of the data at which it is growing. According to IDC [6] report, the volume of data will reach to 40 Zeta bytes by 2020 and increase of 400 times by now. The volume of data is growing rapidly, due to several applications of business, social, web and scientific explorations.

- **Velocity:** The speed at which data is increasing thus demanding analysis of streaming data. The velocity is due to growing speed of business intelligence applications such as trading, transaction of telecom and banking domain, growing number of internet connections with the increased usage of internet, growing number of sensor networks and wearable sensors.

- **Variety:** It depicts different forms of data to use for analysis such as structured like relational databases, semi structured like XML and unstructured like video, text.

- **Veracity:** Veracity is concerned with uncertainty or inaccuracy of the data. In many cases the data will be inaccurate hence filtering and selecting the data which is actually needed is really a cumbersome activity. A lot of statistical and analytical process has to go for data cleansing for choosing intrinsic data for decision making.

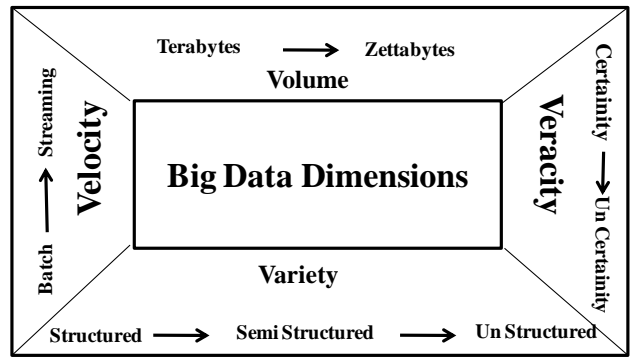

**Figure 5. Data Dimensions 4V's.**

**ii)** Analytics – Big Data techniques

analytics is the process of analyzing the data using statistical models, data mining techniques and computing technologies. It combines the traditional analysis techniques and mathematical models to derive information. Analytics and analysis performs the same function, however, analytics is the application of science to analysis. Big Data Analytics refers to a set of procedures and statistical models to extract the information from a large variety of data sets. A few major Big Data Analytics application areas are discussed below.

- **Text analytics:** The process [45] of deriving information from text sources. The text sources forms of semi-structured data that include web materials, blogs and social media postings (such as tweets). The technology within text analytics comes from fundamental fields including linguistics, statistics and machine learning. In general, modern text analytics uses statistical models, coupled with linguistics theories, to capture patterns in human languages such that machines can understand the meaning of texts and perform various text analytics tasks. Text mining in the area of sentiment analysis helps organizations uncover sentiments to improve their customer relationship management.

- **In Memory analytics:** In memory analytics [26] is the process which ingests the large amounts of data from a variety of sources directly into the system memory for efficient query and calculation performance. In-memory analytics is an approach for querying data when it resides in a computer's random access memory (RAM), as opposed of querying data stored on physical disks. This results in vastly shortened query response times, allowing business intelligence (BI) applications to support faster business decisions.

- **Predictive analysis:** Predictive analysis [46] is the process of predicting future or unknown events with the help of statistics, modeling, machine learning and data mining by analyzing current and historical facts.

- **Graph analytics:** Graph analytics [44] studies the behavior analysis of various connected components, especially useful in social networking web sites to find the weak or strong groups.

### iii) *Technologies*

Big Data technologies are majorly classified into three parts viz. i) file system – effective way of organizing the data ii) computing frameworks, and iii) tools for analytics as described below.

**a. *File System:*** File system is responsible for the organization, storage, naming, sharing, and protection of files. Big Data file management is similar to distributed file system, however the read/write performance, simultaneous data access, on demand file system creation, efficient techniques for file synchronizations would be major challenges for design and implementation. The goals in designing the Big Data file systems should include certain degree of transparency as mentioned below.

- Distributed access and location transparency: Unified directory services, clients are unaware that files are distributed and can access them in the same way local files are accessed. Consistent name space encompassing local as well as remote files without any location information.
- Failure handling: The application programs and the client should operate even with the few components failures in the system. This can be achieved with some level of replication and redundancy.
- Heterogeneity: File service should be provided across different hardware and operating system platforms.
- Support fine-grained distribution of data: To optimize performance, we may wish to locate individual objects near the processes that use them.
- Tolerance for network partitioning: The entire network or certain segments of it may be unavailable to a client during certain periods (e.g. disconnected operation of a laptop). The file system should be tolerant enough to handle the situations and applies the appropriate synchronization mechanisms.

**b.** Open source frameworks: Big Data computing frameworks which are based open source frameworks are described as below.

- Apache Hadoop [7]: An open source reliable, scalable and distributed computing platform. It offers a software library and framework that allows distributed processing of large scale distributed processing of large data sets across clusters of computers using simple programming models.
- Spark [27]: Apache Spark is a fast and general engine for large scale data processing. This covers Shark SQL, Spark Streaming, MLib machine learning and Graphx graph analytics tools. Spark can run on Hadoop YARN [7] cluster manager, and can read any existing Hadoop data.
- Storm [28]: distributed real time stream oriented computing for real time analytics, online machine learning, continuous computation, distributed RPC, ETL etc. Storm topology consumes streams of data and processes those streams in arbitrarily complex ways, repartitioning the streams between each stage of the computation.
- S4 [29]: Platform for processing continuous data streams. S4 is designed specifically for managing data streams. S4 apps are designed combining steams and processing elements in real time.

**c. Commercial frameworks**

Google offers Big Query [30] to operate on Google Big Tables [31]. Amazon supports Big Data through Hadoop cluster and also NoSQL support of columnar database using Amazon DynamoDB [32]. Amazon Elastic MapReduce (EMR) [33] is a managed Hadoop framework that makes it easy, fast, and cost-effective to process vast amounts of data across dynamically scalable Amazon EC2 instances. Windows offers HDInsight [34] service an implementation of Hadoop that runs on the Microsoft Azure platform. RackSpace [35] offers Horton Hadoop framework on Openstack platform, Aneka [26] offers .NET based desktop MapReduce platform, and other enterprise frameworks based on open source Hadoop are Horton [36] and Cloudera [37] .

### d. Tools

The brief descriptions of the several Big Data tools are described below.

- **Key-Value Stores:** Key-value pair (KVP) tables are used to provide persistence management for many NoSQL technologies. The concept is; the table has two columns- one is the key; the other is the value. The value could be a single value or a data block containing many values, the format of which is determined by program code. KVP tables may use indexing, has tables or sparse arrays to provide rapid retrieval and insertion capability, depending on the need for fast look up, fast insertion or efficient storage. KVP tables are best applied to simple data structures and on the Hadoop Map Reduce environment. Examples of key-value data stores are Amazon's Dynamo [32] and Oracle's Berkeley DB [36].
- **Document Oriented Database:** A document oriented database is a database designed for storing, retrieving and managing document oriented or semi-structured data. The central concept of a document oriented database is the notion of a document where the contents within the document are encapsulated or encoded in some standard format such as JavaScript Object Notation (JSON), Binary JavaScript Object Notation (BJSON) or XML. Examples of these databases are Apache's CouchDB [40] and 10gen's MongoDB [41].
- **Column family/big table database**: Instead storing key-values individually, they are grouped to create the composite data, each column containing the corresponding row number as key and the data as value. This type of storage is useful for streaming data such as web logs, time series data coming from several devices, sensors etc. the examples are HBase [42], and Google Big Table [31]
- **Graph database:** A graph database uses graph structures similar to nodes, edges and properties for data storing and semantic query on the data. In a graph database, every entity contains direct pointers to its adjacent element and index look-ups are not required. A graph database is useful when large-scale multi-level relationship traversals are common and desirable for processing complex many-to-many connections such as social networks. A graph may be captured by a table store, which supports recursive joins such as Big Table and Cassandra. Examples of graph databases include Infinite Graph [43] from Objectivity and the Neo4j open source graph database [44].

iv) ***Programming Models:*** various programming models like data intensive, stream computing, batch processing, high performance/throughput, query processing, column oriented data processing described below.

- MapReduce: data intensive programming model, with high level programming constructs for Map and Reduce functions in the cluster of distributed compute and storage nodes. Map function performs filtering and sorting, whereas Reduce function aggregates map output to generate the final result. Map

Reduce programming is a type of recursive programming model to operate the similar logic on multiple distributed data sets. The examples are Hadoop Map Reduce [47] , Apache Spark [27] , Aneka MapReduce [17] .

- Thread/Task Data Intensive Models: Thread programming models are used for high performance applications, the computing logic demands more computing elements or high end cores for processing within to meet the application deadlines and task programming models are used for workflow programming models, e.g. Aneka [17].

- Machine learning tools: new generation of machine learning tools for decision making. Few tools available are Hadoop Mahout [48].

- Big Query Languages: new generation of Query Languages, examples are Google Big Query [30]. Web Log mining is the study of the data available in the web. This involves searching for the texts, words and their occurrences. One example for web log mining is searching for the words and their frequencies by Google Big Query Data Analytics [30] uses Google Big Query platform to run on the Google cloud infrastructure.

Big Data computing majorly need to address two types of scheduling mechanisms; Query scheduler and Data Aware Scheduling. Query schedulers addresses several mechanisms for querying the data managed by Big Data systems, Data Intensive schedulers addresses several computing mechanisms, examples include Capacity scheduler [49], and Fair scheduler [50].

*v) Big Data Security*

Big Data project can uncover tremendous value for an enterprise, by revealing customer buying habits, detecting or preventing fraud, or monitoring real time events. However, a poorly run Big Data project can be a security and compliance nightmare, leading to data breaches. Big Data must be protected, to ensure that only the right people have appropriate access to it. Big Data security addresses several mechanisms for large scale high volume rapidly growing varied forms of data, analytics and large scale compute infrastructure. As, the data volumes and compute infrastructures are very large, traditional methods of computing and data security mechanisms, which are tailored for securing small scale data and infrastructure, are inadequate. Also, the use of large scale cloud infrastructures, with a diversity of software platforms, spread across large networks of computers, also increases the attacks. The onion model of defense for Big Data Security is depicted in Figure 6 and the several elements are described below.

- Distributed computing infrastructure: Mechanisms for providing security while data is analyzed over a multiple distributed systems. Big Data setup would be either confined to an enterprise or could be a large collection of several enterprises, social and scientific collection of disparate sources distributed system. Privacy, Security and Confidentiality - Not revealing private and confidential information to unauthorized users. For example, in a mailing system, secrecy is concerned about preventing the users from finding out the passwords of other users.

- Large scale distributed data:  privacy preserving mechanisms, encryption techniques for the data stored on large scale distributed systems,  role based access and control mechanisms, and security of column, document, key-value and graph data models to be evolved. In order to maintain fast access for the data, NoSQL databases come with little built-in security, due to their BASE (Basically Available, Soft state, Eventually consistent) properties; rather than requiring consistency after every transaction, the data base just needs to eventually reach a consistent state.

- Analytics security: developing frameworks which are secured, that allows organizations for publish and use the analytics securely based on several authentication mechanism such as one time passwords(OTP), multilevel authentications and role based access mechanisms.
- Users privacy and security: confidentiality, integrity and authentication mechanisms to validate the users.

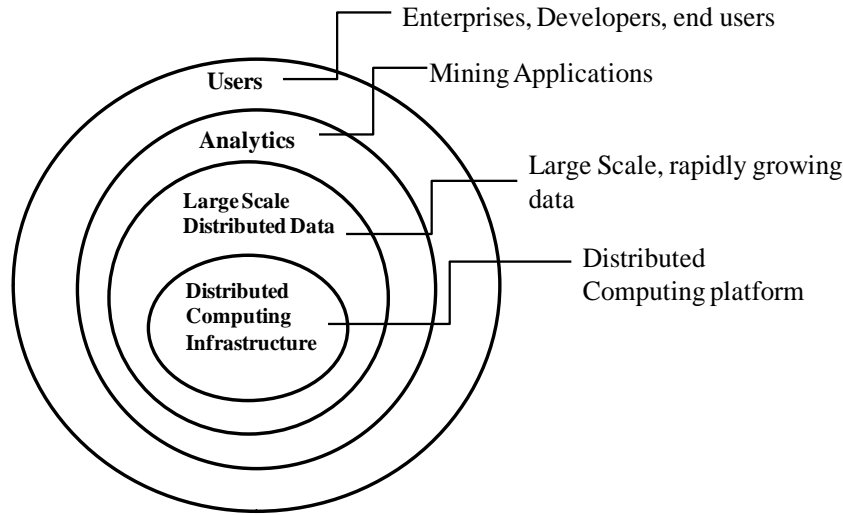

**Figure 6. Big Data Security Onion Model of Defense.**

## 4 Big Data in Clouds: An Integrated Big Data and Cloud Platform

Big Data in clouds is a new generation data intensive platform for quickly building the analytics and deploying over a elastically scalable infrastructure. Based on the services rendered to the end users, these are broadly classified into four types as described below.

• **Public Big Data clouds:** Large scale data organization and processing over the elastically scalable clouds infrastructure. The resources are served over internet as pay-as-go computing models. The examples include Amazon Big Data computing in clouds [33] , Windows Azure HDInsight [34] , Rack space Cloudera Hadoop [35] [37] , and Google cloud platform of Big Data computing [30] .

• **Private Big Data clouds:** deployment of Big Data platform within the enterprise over a virtualized infrastructure, with a greater control and privacy to the single organization.

• **Hybrid Big Data clouds:** Federation of public and private Big Data clouds for scalability, disaster recovery and high availability. In this deployment, the private tasks can be migrated to the public infrastructure during peak workloads.

• **Big Data access networks and computing platform:** Integrated platform of data, computing and analytics delivered as services by multiple distinct providers.

Big Data computing in Clouds aka **"Big Data Clouds"** is data intensive analytics platform of large scale, distributed compute and storage infrastructures. The features of Big Data Clouds are i) large scale distributed compute and data storages: wide range of computing facilities with seamless access to scalable storage repositories and data services ii) information defined data storage: meta data based data access instead of path and filenames iii) distributed virtual file system: File system could be dynamically created and mapped to the computing cluster iv) seamless access of computing and data: Transparent access to large scale data and compute resources iv) dynamic selection of data containers and compute resources: able to handle dynamic creation of  virtual machines and able to access large scale distributed data sources increasing the data location proximity v) High performance data and computation: compute and data should be high performance driven vi) Multi dimension data handling: Support  for several forms of data with necessary tools for processing vii) Analytics platform services: able to develop, deploy and usage Analytics over the environment viii) High availability of computing and data: replication mechanisms both for computing and data ix) platform for data intensive computing: support for both traditional and emerging data intensive computing models and scalable deployment and execution of applications.

Figure 7 depicts Integrated Cloud and Big Data access networks on Cloud infrastructure for analytics development. The content from several sources like social media, web logs, scientific studies, sensor networks, business transactions etc. are growing rapidly. Deriving useful information for decision making from such large data, fusing the information from several sources would be a challenging task. The elements of Big Data access network are; Data services, Big Data computing platform, Data scientist and Computing Cloud described below.

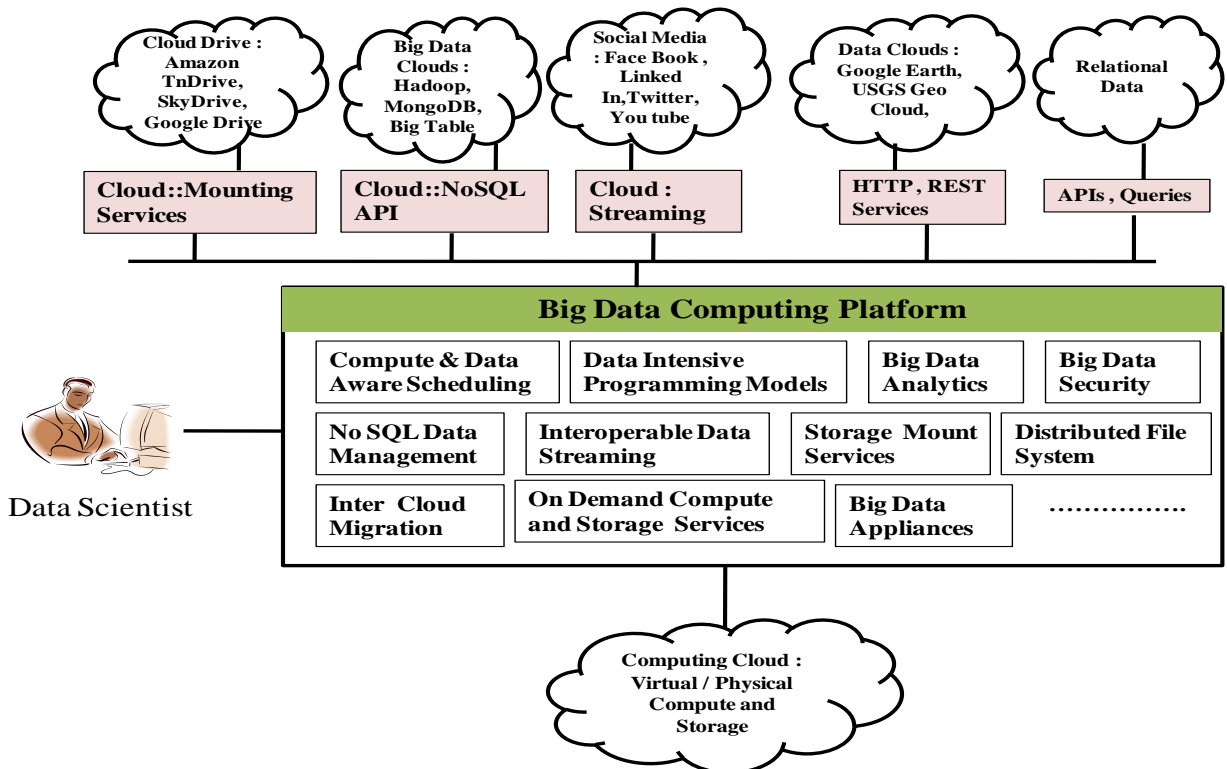

**Figure 7. Integrated Cloud and Big Data Compute Network.**

• **Data and Platform Services**: several providers, those who provide services for accessing both data and platform services for computing on the data, for example; Google Data APIs (GData) [51] provide protocols for reading and writing data on the web for several services like Content API for shopping, Google Analytics, Spreadsheets and YouTube.

- **Big Data Computing Platform:** Platform for managing the various data sources including data management, access, programming models, schedulers, security etc. The platform includes various tools for accessing other data platforms using streaming, web services, APIs. Other data platforms includes data services from relational data stores, google data, social networking etc.
- **Data scientist:** analytics developers having access to the computing platform.
- **Computing Cloud:** computing infrastructure from private/public/hybrid clouds.

## 4.1 *Big Data Clouds for the Enterprise*

Big Data clouds enable enterprises to save money, grow revenue and achieve many other business objectives in any vertical by quickly building their Big Data databases and writing analytics for mining the information. The benefits of Big Data clouds for the enterprises are mentioned below.

- **Build new applications**: Big Data clouds would allow enterprises to collect billions of real-time data points on its products, resources or customers and then repackage that instantaneously to optimize customer experience or resource utilization.
- **Improve the effectiveness and minimize the cost**: Big Data clouds offer services and pay as go consumption model similar to cloud services. This pricing model would effectively reduce both the cost of the applications development by minimizing the cost of development tools.
- **Realize new sources of information and build applications to gain competitive advantage:** The information could be quickly fused from several Big Data databases and rapidly build applications for several platforms like hand held and mobile devices.
- **Increase in customer loyalty:** Increase in the amount of data sharing within the organization and the speed with which it is updated allows businesses and other organizations to more rapidly and accurately respond to customer demand.

## 4.2 *Elements of Big Data Cloud*

Big Data and traditional data ware housing mechanisms differ with each other in several ways like large scale data organization, and querying followed by platforms and tools to the Data Scientists for analytics development. In this section, we describe elements of Big Data Cloud as shown in Figure 8.

i) **Big Data Infrastructure Services (BDIS):** This layer offers core services such as compute, storage and Data Services for Big Data Computing as described below.

a) Basic storage service: provide basis services for data delivery which is organized either on physical or virtual infrastructure, and support various operations like create, delete, modify, and update with a unified data model supporting various types of data.

b) Data organization and access service: Data organization provides management and location of data resources for all kinds of data, and selection, query transformation, aggregation and representation of query results, and semantic querying for the selecting the data of interest.

c) Processing service: Mechanism to access the data of interest, transferring to the compute node, efficient scheduling mechanism to process the data, programming methodologies, various tools and techniques to handle the variety of data formats.

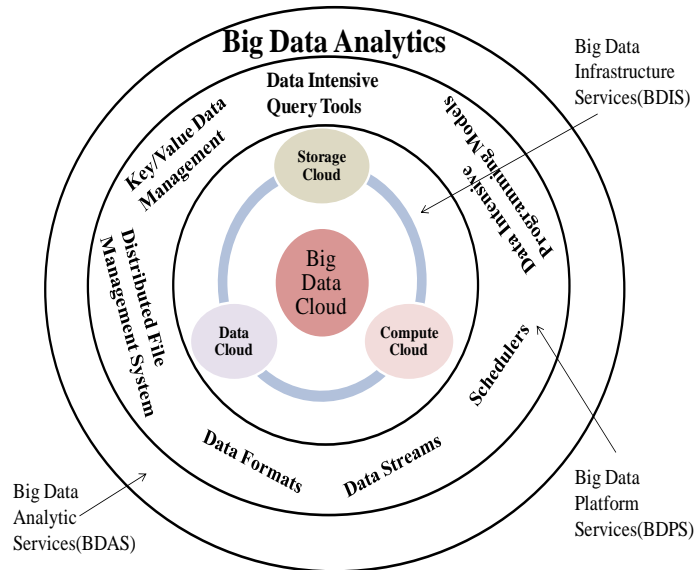

**Figure 8. Big Data Cloud Components.**

The elements of BDIS are described below.

- Computing Clouds: On Demand provisioning of compute resources, which could expand or shrink based on the analytics requirements.
- Storage Cloud: Large volume of storage offered over the network. The storages offered include; file system, Block Storages and Object Based Storage. Storage clouds offer to create file system of choice and also elastically scalable. Storage Clouds can be accessed based on the pricing models which are usually based on data volumes, transactions/data transfer. The several services offered by Storage Clouds are Raw, Block and Object based storages
- Data Clouds: Data Clouds are similar to Storage Clouds, however, unlike storage space delivery, they offer data as a service. Data Clouds offer tools and techniques to publish the data, tag the data, discovery the data and process the data of interest. Data Clouds operate on domain specific data leveraging the Storage Clouds to serve Data as a Service based on four step of "Standard Scientific Model" [40] such as data collection, analysis, Analyzed reports and long term preservation of the data.

ii) **Big Data Platform Services (BDPS):** This layer offers schedulers, Query mechanisms for Data retrieval, Data Intensive Programming models to address several Big Data Analytic problems.

iii) **Big Data Analytics Services (BDAS):** Big Data Analytics as Services over Big Data Cloud infrastructure. The services would be offered to enterprises based on SLAs meeting QoS parameters.

### 4.3 *Big Data Clouds Layered Architecture*

The architecture of Big Data computing in clouds is represented as four layered model as shown in Figure 9. The Cloud Infrastructure layer handles the elastic scalable computing, storage and networking infrastructure. The Big Data fabric layer; addresses the several tools for data management, access and aggregation. The third layer is the platform layer which addresses the tools and technologies for data access and processing, programming environments for designing the analytics and scheduling models for execution etc. the top layer is the Big Data analytics, focused on analytics usage, and publishing standards to offer them as services. The functional description of each of the layers is described below.

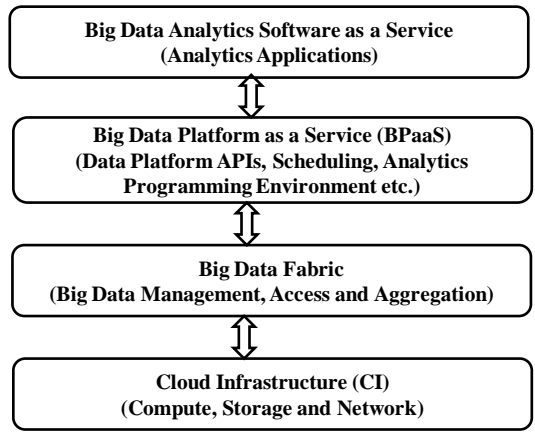

**Figure 9. Big Data Cloud Reference Architecture.**

**a) Cloud infrastructure (CI):** Large scale management of dynamic and elastic scalable large infrastructure of compute and storage resources as services. Virtualization technologies are used for on demand provisioning of the resources based on SLAs and QoS parameters. The services rendered by this layer are i) large scale elastic infrastructure to setup Big Data platform on demand ii) dynamic creation of virtual machines iii) large scale data management for File/Block/Objected based storages on demand iv) ability to move the data in seamless across the storage repositories, and v) able to create the virtual machines and auto mount the file system with the compute node.

**b) Big Data fabric:** This layer addresses tools and APIs through which storage, compute and application services can be accessed. This layer offer interoperable protocol APIs to connect multiple cloud infrastructures standards specified [52].

**c) Big Data Platform as a Service (BPaaS):** Core layer offers several platform services to work with storage/data, and computing services based on SLAs and QoS. This layer consists of middleware management tools such as schedulers, data management tools such as NoSQL tools, and data intensive programming models for data processing. This layer would mainly focus on development of tools and SDKs which are essential for the design of Analytics.

**d) Big Data Analytics:** Big Data analytics offered as services, where, users could quickly work on analytics without investing on infrastructure and pay only for the resources consumed. This layer organizes the repository of software appliances and quickly deploys on the infrastructure and delivers the end results to the users, the pricing would be computed based on the usage, quality of service provided etc.

### 4.4 *Layered components*

The layered architecture, sub layers and the components in each layer are shown in Figure 10 and Table 2 describes the layered architecture and their corresponding mapping of the reference architecture.

**Table 2. Layers mapped to reference architecture**

| S.no | Layer name | Sub layers | Reference layer of architecture |
|------|------------|------------|--------------------------------|
| 1 | Infrastructure | Resource and Interface layers | Cloud Infrastructure (CI), Big Data fabric |
| 2 | Big Data Platform | Foundation, Runtime, Programming modeling layer, SDK | Big Data platform as a Service |
| 3 | Applications | Analytics, Big Data services | Big Data Analytics software as a Service |

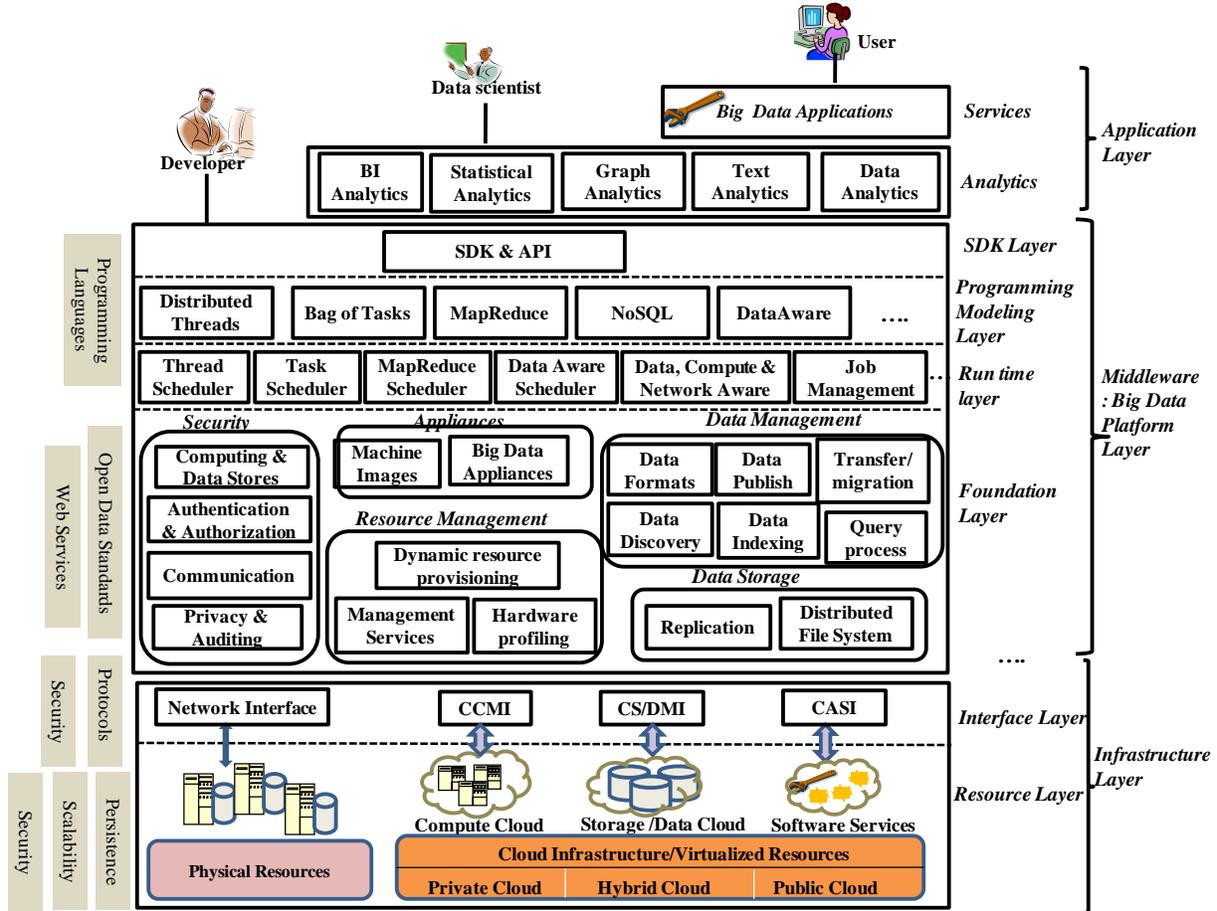

**Figure 10. Big Data Cloud Layered Components.**

## a) Infrastructure layer

This layer provides services for effective management and delivery of the computing elements, storage, data and networking infrastructure. This layer is further classified into two sub layers; resource and interface layers. Resource layer facilitates compute, storage and data services either on physical or virtual environments. Physical environment is similar to data centers without virtualization enforced and is similar to cluster setup in the local network. In the case of virtual environments, it could be a private/public/hybrid cloud provider who offer services based on the consumption. The functioning of resource layers over physical and virtual environments is similar, however, virtual environments offer high utilization of the resources, on demand resource provisioning and highly scalable, however, endure performance degradations due to enforced virtualization technologies. Below, services offered by resource and interface layers are described in brief.

**i. Resource layer:** Resource layer handles both physical and cloud resources as discussed below.

**a) Physical resources:** non-virtualized compute and storage resources delivered via local data centers or in-house available. The resources may be accessed via standard protocol and networking interfaces.

b) **Virtualized/Cloud resources:** The resources are delivered by several cloud providers like compute, storage and application clouds. Compute clouds offers several scalable machine instances on demand, storage / data clouds offers either storage repositories or data online, and sometimes both. Software services are similar to applications offered as services over the cloud. The cloud infrastructure may be private, public, or both. However, the access mechanisms and security implementations will differ depending on types of clouds which were chosen for the setup. Below, we illustrate the functions of compute, storage/data cloud and software services.

**i) Compute cloud:** large pool of compute machine instances to serve the demands. Compute machines could be created at run time and the data needed for analytics purpose may be made available dynamically.

**ii) Storage/Data cloud:** Storage clouds offers a pool of the storage space where in files required for analytics could be placed. However, Data Cloud offers the storage space along with the Data necessary for compute. Such data or storage could be offered as Block Storage or Object Storage.

**iii) Service cloud:** pre built analytic tools, provisioned on demand for quickly accessing the underlying data and computing resources.

**ii. Interface Layer**

Interface layer facilitate open standards, protocols based on web and interoperable services. The major challenges include; interoperability between heterogeneous hardware and storage infrastructure, and migration/access across various cloud providers. Interface layer offers standard interfaces [52] to access compute resources, storage resources and application services. This layer could be classified into four components based on the services rendered to the foundation layer, such as networking interface protocols, cloud compute management interface (CCMI), cloud Storage/Data management interface (CS/DMI) and Cloud application services interfaces (CASI). The detailed description for each of the components is given below.

- **Network Interface:** This interface allows several physical devices access through standard networking interfaces and protocols. This includes accessing the compute instances via terminal services or web consoles. The storage devices can be mounted to the local compute machine instances or access via separate networks such as NFS protocols.

- **CCMI (Cloud Compute Management Interface):** Interoperable functional interfaces for on demand creation and management of virtual machines of several public cloud providers.

- **CS/DMI (Cloud Storage / Data Management Interface):** A functional interface that applications will use to create, retrieve, update and delete data elements from the Cloud. As part of this interface, the client will be able to discover the capabilities of the cloud storage offering and use this interface to manage container and the data that is placed in them. In addition, Metadata can be set on container and their contained data elements through this interface. This interface is also used by administrative and management applications to manage container, accounts, security access and monitoring/billing information, even for storage that is accessible by other protocols. The capabilities of the underlying storage and data services are exposed so that clients can understand the offering. Various CDMI interfaces are

- o **Amazon S3**: Amazon S3 stands for Simple Storage Service, stores the data objects within the buckets, comprised of a file and optionally any metadata that describes that file. To store an object in Amazon S3, the file can be uploaded to the bucket and permissions can be set on the object as well as on the metadata. Buckets are the containers and there can be more than one bucket.

- o **Open Stack Swift**: Object Based Data Storage system exposes the storage via REST API and stores a large amount of unstructured data at low cost.

- **CASI (Cloud Application Services Interface):** set of Web Services that exposes the published applications through standard web protocols. This also involves application virtualization methodologies to serve only the needed applications as services from the cloud providers.

**b) Big Data Platform layer**

This is a middleware layer that is further categorized into four sub layers based on the functionality, they are, foundation layer, runtime layer, programming modeling layer and software development kit (SDK) layer. The Foundation layer offers mechanisms for resource management, data storage, data management, security and virtual appliance. The Runtime layer addresses several scheduling mechanisms and job management mechanisms. The Programming Modeling layer employs several programming standards; the SDK layer offers Application Programming Interfaces (APIs) for programming in several languages. The detailed description of the layers is given below.

- **Foundation layer**: This is the core part of the middle ware layer, which interfaces with the resource layer. This layer mainly classified into components such as resource management, data management, appliances, data storage and security.

  - o **Resource management:** Resource management consists of the following components.
    - **Management services**: The services for managing the underlying physical resources. These can be middleware services to track of the available resources. The management services include applications to monitor the resource utilizations for data and compute such as compute resources availability, storage availability etc.
    - **Hardware profiling**: The information services to retrieve the information regarding the available resources such as RAM, Network Bandwidth, compute load etc.
    - **Dynamic resource provisioning**: Facilitates the resources at run time from the virtualized resources.

  - o **Data management:** This mainly deals with data formats, discovery, and publishing mechanisms.
    - **Data Formats:** Data formats service provides to store the data in various types of forms which include structured, unstructured and semi structured. Search mechanisms services offer various query mechanisms to search for the data of interest, Sharing allows various access privileges.

▪ **Data transfer/migration:** The mechanisms either pull or push the data for processing, automatic syncing of the files to the Big Data systems. It also contains tools that are necessary for migration of the existing structured/unstructured data to cloud Big Data workloads.

▪ **Data discovery mechanisms:** Several mechanisms to find the location of the data. This can be performed with the query mechanisms or searching the metadata contents. Dedicated discovery mechanisms for specific communities need to be evolved. Technologies for data discovery might include visualization, structural query mechanisms, semantic query etc.

▪ **Data publication mechanisms:** Several mechanisms of domain specific data publication and retrieval mechanisms.

▪ **Data indexing:** Indexing mechanisms are needed to speed up the process of accessing the data. Several Data Indexing mechanisms need to be explored for data redundancy, replication.

▪ **Query process:** Several query processing methods for quickly analyzing the large scale distributed data of both structure and unstructured.

o **Appliances:** self-configuration, appliances eliminates the time consuming efforts of choosing and configuring hardware, determining the proper software components, integrating and tuning the overall configuration.

▪ **Machine image**: The repository of machine instances for creating the systems on demand, virtual machine manager for automated management of the systems and machine image instance, which are pre built machine templates.

▪ **Big Data appliances**: The repository and management of Big Data Appliances which are specific for the domain specific analytics development.

o **Data storage**

• **Replication procedures**: Several replication procedures for duplicating the data onto multiple storage repositories for data redundancy, high availability and high performance data transfers. Instead, repositories confined to a single location, the data would get replicated to multiple geographical locations. Some of the issues addressed Replication [18] techniques specific to Big Data are - i) providing extremely rapid access to data from multiple sources, even in a mixed workload ii) Reducing the drag on multi-way joins for complex queries iii) Accelerating reporting for faster analysis, review and decision making, and iv) backup and disaster recovery techniques. For example Tervela [53] accelerates Big Data Replication by efficiently duplicating to multiple sites with ease through one of two methods such as changed data capture or parallel replication.

▪ **Distributed File System:** File system which stores the data onto multiple distributed storage repositories. The file system maintains the indexing of the data and offers various logical views of the entire data available in the system.

o **Big Data security**: Privacy preserving, auditing, Role based access mechanisms for providing security to the data for both data at rest and while in transit.

• **Runtime layer**: This layer is concerned about workload handling with the support of several scheduling mechanisms. Examples include Thread, Task, Map Reduce, Data and network aware scheduling, batch job management based on the type of computation needed. Below, we briefly, discuss the functions and characteristics of several types of schedulers.

- o **Thread scheduler**: Thread Scheduling exploits the available cores/processors effectively by utilizing either local system or remote system resources. Local threads execution could use shared memory, however, for remote execution; objects migration would take place. Thread scheduling, is applicable for problems that are recursive, multiple data streams but applied on a single instruction. Thread scheduler determines the best resources for running the several spawn independent threads on available resources/cores. Thread Scheduling addresses high performance computing problems.
- o **Task scheduler:** Distributed processing of tasks on several computing nodes. Task Scheduling solves the high throughput problems by determining the best available resources for execution.
- o **Data Aware scheduler:** Jobs execution, knowing the best available storage locations for execution or transfer the data to compute nodes from the best available store repositories. The process could depend on computing the best replicated site that minimizes the compute time. This may apply Data Parallelism to pull/push the data to the compute nodes. Map Reduce is an example of Data Aware Scheduler.
- o **Map Reduce scheduler:** Type of Data Aware Scheduling which maps the compute process to the data nodes. After the completion of the process, the results are consolidated onto a single node for final result.
- o **Data Compute and Bandwidth aware Scheduler:** Considering compute, network and data to solve the data intensive scheduling. The techniques applied for this type of scheduling are
  - i. **Parallel Data Extractor**: Parallel Data Extractor is the high performance Data Transfer module. It enables extraction of transfer of data from storage clouds to the compute node. This module pulls the data from the storage repositories by establishing multiple parallel lines between storage clouds to compute resources. Parallel Data Extractor identifies possible data storage resources and identifies the amount of data to be pulled from each of the storage repositories.
  - ii. **Scheduler**: The scheduler which effectively maps a set of jobs to the computing nodes, the scheduling would depend on heuristic approaches. Big Data schedulers could pick up the best computing nodes or may quickly clone the virtual machines and perform the computation by applying effective data aware scheduling techniques.
  - iii. **Job Management:** Management tools for monitoring the job executions.

- **Programming modeling Layer**: Several programming models to solve the Big Data Problems. This may include; coarse/fain grain programming models for thread, tasks and Data Intensive. It also addresses data handling and query programming models for NoSQL databases.
- **SDK layer:** The programming APIs to solve Big Data Problems. This could be Java, C, C++, and C# based APIs.

- **Application layer:** Application layer provides various statistical, deterministic, probabilistic, machine learning techniques for developing the domain specific analytics tools. This layer offers SDKs; APIs & Tools for the analytics development and also responsible for several management Interfaces development for monitoring the Big Data environments. The several analytics include; Statistical Models, Graph Analytics, Business Analytics, Text Analytics, and Data Analytics.

**c) Users**: The several stake holders of the systems like i) Developers: Big Data general purpose application designer ii) Data Scientist: Data Analysts who design the Analytics applications. This could be Business Analytics, Scientific Explorations etc., iii) End Users: Analytics users of the system.

## 5 Gap Analysis and Future Directions

Big Data research area is broadly classified into four major segments denoted as 4D's i.e. Depository, Devise, Domain and Determine as shown in Figure 11. Depository deals about the storage technologies, devise is about working on new platforms and programming models, Domain is latest trends platforms and tools specific to the various engineering domains, and finally determine is about working on the analytics for mining and information extraction. In this section, we discuss sub elements in each segment described as below.

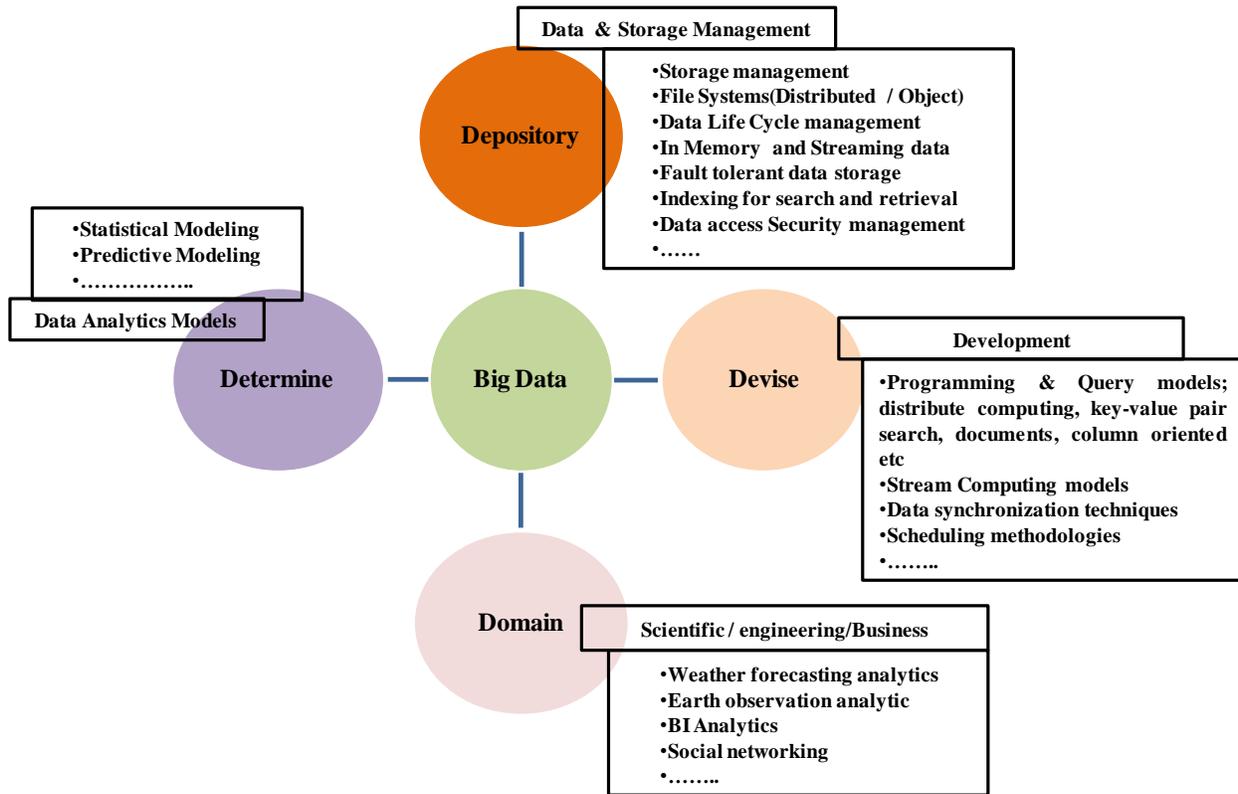

**Figure 11. Big Data Research segments.**

**5.1 Depository:** This segment deals the long term persistent storage and retrieval of both structured and unstructured data of geographical dispersed locations. The several research areas include- migrating from the traditional storage like storage area networks, and network attached storage to the container based object storage systems. This would eliminate hierarchical file structure handling (like directories and subdirectories), eliminate the need of web servers, and load balancers as accessed via HTTP based REST services, facilitates location transparency with unique object identification, high available and fault tolerant storages dispersed across distributed locations described in Table 3.

**Table 3. Depository: gap analysis and future directions.**

| Key element | Gap Analysis and future directions |
|---|---|
| **1. Storage and network** | • The unified storage systems with the combination of three layers traditional storage architectures like file system, volume manager and data protection with wide range of industry standard protocols, including network file system (NFS), Server Message Block (SMB), Hypertext Transfer Protocol (HTTP), File Transfer Protocol (FTP), REST-based Object access etc. Converged I/O protocols like InfiniBand need to be investigated for large volume I/O operations and analytics.<br>• Write Once and Read Many (WORM) object storage technologies for both long term preservation and mining. The indexing techniques need to be explored either attached with the file object or as separate Meta file using NoSQL data structures.<br>• New WAN-based protocols needs to be investigated as the traditional WAN-based transport methods cannot move terabytes of data. These transport methods use effective bandwidth and achieve transfer speeds.<br>• Storage protocols need to be explored for on disk data encryption, privacy preserving, and query on encrypted data mechanisms with reasonable good speeds. |
| **2. File system** | • Conventional file systems have constraints in name space handling as they are assigned by the operating system. This would difficult the searching process while the data is mostly unstructured. Hence, new storage technologies, such as object storage services would allow the files to assign user defined metadata separating it from the operating systems.<br>• Information defined Data Storage complements the existing distributed file system to derive the value out of the data by its content and meaning but not just with names.<br>• File system migration tools of the traditional file systems to object based storage systems to be evolved.<br>• High performance parallel data transfer with data distribution, replication and redundant mechanisms. This could be achieved by replicating the big data file objects on multiple distributed storage systems and creating the clustered indexes.<br>• Data migration tools need to be addressed for moving the on premise big data files to the cloud based big data systems which could be relational data or unstructured like documents, texts, videos, audios etc. |
| **3. Data Life Cycle Management (DLM)** | • Data life cycle addresses the data management issues at several phases of data creation, usage, sharing, storing and eventually archiving or disposing automatically based on policies defined within the management. BDLM (Big Data Life Cycle Management) systems need to be developed incorporating several user defined policies for data organization and minimize the storage resource costs. For e.g., the policy could be data aging, addresses issues related to the data obsolete, something like, deleting the age old objects. Another policy could be the version management, holding only the recent versions of each object in |

| | a bucket with a versioning enabled. |
| | • This N-tier storage architectures could organize the data which is mostly used in lower tiers (tier1) such as Flash/SSD and migrate the data moving down to other tiers such as flat disks, capacity disks to tier N such as Cloud storage for backup/archival.. |
| **4.      In memory computing (IMC) systems** | In-memory computing systems with object storages file systems for effective computing methods and long term preserving need to be evolved. It also enables the querying the data based on the metadata from the cloud storage pools and perform the object based data storage, query and object based data analysis. |
| **5.High Availability(HA) and Fault Tolerant(FT) Data Storage** | HA systems offer storage providing multiple internal components and multiple access points to storage resources. In other words, the system has a second critical component or path to data available in case something fails. This availability or single point of failure doesn't eliminate downtime. Instead, it minimizes it by restoring services behind the scenes, in most cases before the user notices failures. |
| **6.Indexing for search and retrieval** | Indexing multidimensional data and enabling object based retrieval mechanisms instead of set based needs to be developed for efficient query processing. |
| **7.Data access and Security** | Data access and security mechanism in Big Data Clouds needs to be developed which set policies enabling which users get access to which original data, with protection of sensitive data that maintains usable, realistic values for accurate analytics and modeling on data. |

## 5.2 Devise–Big Data Platform Services

This segment focused on the design of platforms and programming models in distributed computing, in-memory computing, stream computing, query languages for assorted data such as key-value data stores, column oriented data, document databases, content synchronization techniques, and scheduling methodologies etc. Recently, several Big Data computing platforms are emerged such as Hadoop, Spark, Amazon EMR, Dryad, HDInsight, Aneka and Map Reduce [54], [55] have become a de facto standard of big data applications over a large scale cluster of computing nodes. However, Map Reduce, has limitations both from theoretical perspective [56], [57] and empirically by exploring classes of algorithms that cannot be efficiently implemented [58], [59], [60], [61]several limitations of the Map Reduce Model over Hadoop File System are described below.

### Table 4. Device: gap analysis and future directions

| Key element | Gap Analysis and future directions |
|---|---|
| 1.**Programming models** | • Map Reduce programming addressing iterative and non iterative models, with storage coupled with computing nodes and as separate services need to be investigated. Current MapReduce models are iterative in nature [62], hence the problems like Page Ranking Iterative Graph Algorithms, Gradient Descent cannot be addressed. Also, the current models of MapReduce to be modified or extended to address the several problems in engineering and scientific domains like Data Product Generation [63], [64], [65]and DEM [66], [67], as the processing demands large data, hence several data aggregation and processing |
| | scheduling mechanisms to be evolved. |
| | • Debugging tools and profilers for Map Reduce Programming model required to be investigated for Map Reduce programming. Currently, there are batch based without user interaction. |
| | • Domain specific languages need to evolve such data intensive, high performance, IOT programming etc. to solve specific problems in several fields of final services, business sectors, scientific explorations, sensors networks etc. |
| 2.**Unstructured data processing** | • Document, text and graph based data processing mechanisms with better indexing mechanisms to be explored. This could use key value pair mechanisms with schema less data bases and map reduce functions for effectively retrieving the data by processing on the cluster of machines.<br><br>• Unstructured database systems to be explored to bride gap between traditional databases and key value pair databases. These data base systems should work on object notations and perform query on multiple nodes to improve the performance. |
| 3.**Scheduling methodologies** | • Evolutions of new programming models for compute intensive Big Data Problem: programming models with the combination of Thread, Task and Map Reduce need to be devised. The current MapReduce programming model will transfer the compute to the data node. Here, Compute is considered to be a small activity, as compared with Data. This model will not be applicable while Compute is as large as data. Hence new programming models to be exploited.<br><br>• QoS based Resource Management scheduling methods needs to be developed which would work on parameters like time, budget, accuracy etc.<br><br>• HPC programming models such as high speed in memory computing, stream computing need to be worked for HPC Big Data Clouds for scientific applications to address data science problems with accuracy in real time. |
| 4.**Workspace Management** | • Collaborative framework for analytics development to be developed. These frameworks organize the source code, data etc. in sharing mode, and allow the analysts to design and develop the applications in both on premise and cloud based platforms. |

## 5.3 Domain  - Scientific, Engineering and Business

Big Data Analytics extract information from large data for decision making. Examples include, earth observation systems, disaster management study, weather forecasting, simulations, engineering design problems, business intelligence applications. It is also necessary to  evolve several complex applications for monitoring historical data apart from the operational data. The researchers should focus on the following activities to address the domains of data science.

**Table 5. Domain: gap analysis and future directions**

| Key element | Gap Analysis and future directions |
|---|---|
| **1.  Data Management architectures** | • Data management architectures in several domains of geo spatial [68]  [69] , healthcare [70] , social networking and web log mining [71] [72]   still to be |

| | explored. The data gathered, and processed in these fields is unstructured and domain specific, hence indexing architectures, metadata management schemes, query and processing tools specific to the domains still to be investigated. |
|---|---|
| **2. Data visualization models** | • Visualization tools have to be investigated for presenting the large scale data and the analyzed/processed results over dashboards, reports, and charts [73] [13] . |
| **3. Domain specific analytic models** | • Domain specific analytics tools that would pick up the appropriate NoSQL databases necessary for the analytics needs to be explored.<br>• New domain models to be investigated, for migration of the existing in house domain specific analytics to clouds. These models address the data management issues, extract, transformation and load tools for the in house data with effective indexing, processing and tools for analysis.<br>• Analytical models to work on the interested data regions and assigning the score based on the ranking. This could enable the Analytics to pick up the most relevant data for analysis increasing the system throughput. |

## 5.4 Decision – Mining and Determining

Big Data processing is driven by statistical and analytical models to derive information for decision making. Big Data is not just about the data, but the ability to solve business and scientific exploration problems, providing new business opportunities and thoughts. Data Analytic plays a key role in information mining and derives a value out of the data. At present, several analytic systems are evolving, but the majority systems are based on the frameworks which are general purpose tools. Hence, it is essential to address several frameworks in the areas of predictive analysis, behavior analysis, business intelligence etc. In Big Data mining, several open source initiatives tools are becoming popular, as mentioned below. A few research challenges are described below.

• Evolve new architecture for analytics to deal with both historical and real time data at the same time. This could be achieved by organizing the unstructured data as N-tier system with effective indexing and performing the distributed queries and data intensive programming techniques to analyze the historical data and compare with the present data to derive the intrinsic information.

• Statistical significance tools need to be developed to determine maximum like hood (e.g. p-value [73]) for the evidence based on the probabilities and statistics, rather on randomness of data distribution.

• Distributed parallel data mining algorithms and frameworks for unstructured large volume of data need to be investigated, to analyze the data quickly and provide the results summary.

• Time evolving data mining techniques need investigated for the evolving data sets such as words, graph analytics for social networking, behavior analytics, predictive analytics, earth observation geo intelligence solutions, weather forecasting, etc.

• New techniques need to be evolved which could quickly identify the portion of the data needs to be mined rather as a whole to quickly deliver the analysis results.

• Big Data computing in clouds and Cloud based analytics services need to be developed for domain specific applications meeting QoS, SLAs, and budget, deadline constraints.

• Distributed Real-time, predictive and prescriptive analytics tools need to be evolved that could provide the interactivity to the running jobs, apply statistical tools to determine the information and offer the results in the real time [13].

## 6  Summary and Conclusions

Big Data computing is an emerging platform for data analytics to address large scale multidimensional data for knowledge discovery and decision making. In this paper, we have studied, characterized and categorized several aspects of Big Data computing systems. Big Data technology is evolving and changing the present traditional data bases with effective data organization, large computing and data workloads processing with new innovative analytics tools bundled with statistical and machine learning techniques. With the maturity of Cloud computing technologies, Big Data technologies are accelerating in several areas of business, science and engineering to solve data intensive problems. We have enumerated several case studies of Big Data technologies in the areas of health care studies, business intelligence, social networking, and scientific explorations. Further, we focus on illustrating how Big Data databases differ from traditional data base and discuss BASE properties supported by them.

To understand Big Data paradigm, we presented taxonomy of Big Data computing along with discussion on characteristics, technologies, tools, security mechanisms, data organization, scheduling approaches, etc. along with relevant paradigms and technologies. Later we presented under pinning technologies for the evolution of Big Data and discussed how cloud computing technologies would be utilized for infrastructure services delivery for the analytics development. Later, we discussed an emerging Big Data computing platforms over Clouds, Big Data Clouds, an integrated technology from Big Data and Cloud computing, delivering Big Data computing as a service over large scale clouds. The paper also discussed types of Big Data clouds and illustrated Big Data access networks, an emerging data platform services for Big Data analytics.

Further on, we presented a layered architecture, components under each of the layers followed by technologies to be addressed under each of the layers. We then compare some of the existing systems in each of the areas and categorize them based on the tools and services rendered to the users. In doing so, we have gained an insight into the architecture, strategies and practices that are currently followed in Big Data computing. Also, through our characterization and detailed study, we are able to discover some of the short comings and identify gaps in the current architectures and systems.  These represent some of the directions that could be followed in future. Thus, this paper lays down a comprehensive classification framework that not only serves as a tool to understanding this emerging area but also presents a reference to which future efforts can be mapped.

To conclude, Big Data technologies are being adopted widely for information exploitation with the help of new analytics tools and large scale computing infrastructure to process huge variety of multi-dimensional data in several areas ranging from business intelligence to scientific explorations. However, more research needs to be undertaken, in several areas like data organization, decision making, domain specific tools and platform tools to create next generation Big Data infrastructure for enabling users to extract maximum utility out of large the volumes of available information and data.

### Acknowledgements

We thank Rodrigo Calheiros, Nikolay Grozev, Amir Vahid, and Harshit Gupta for their comments and suggestions on improving the quality of paper. This work is partially supported by Future Fellowship project of the Australian Research Council and Melbourne-Chindia Cloud Computing (MC$^3$) Research Network.